\documentstyle[preprint,amstex,aps,eqsecnum,epsfig]{revtex}
\tighten
\def\E{{\cal E}}                            % F
\def\N{{\cal N}}                            % N
\def\T{{\cal T}}                            % T 
\def\V{{\cal V}}                            % V 
\def\a{\alpha}

\def\b{\beta}

\def\l{\lambda}

\def\s{\sigma}

\def\t{\theta}

\def\ZZ{{\Bbb Z }}
\def\bds#1{\boldsymbol{#1}}
\def\lowmp{\lower.11em\hbox{${\scriptstyle\mp}$}}

\def\der#1#2{{d#1\over d#2}}
\def\pdif#1#2{{\partial #1 \over \partial #2}}
\def\braket#1#2{\left\langle #1 \right| \left. #2 \right\rangle}

\def\vev#1{\langle #1 \rangle}
\def\mbare{m^2_{\text{b}}}
\def\lbare{\lambda_{\text{b}}}

\begin{document}
\preprint{ IFUM 650/FT-99 \,\, Bicocca-FT-99-37\bigskip}
%\draft
\title{\bf Out--of--equilibrium dynamics of large$-N$ 
$\phi^4$ QFT in finite volume}
\author{{\bf C. Destri $^{(a,b)}$}  and {\bf E. Manfredini $^{(b)}$\bigskip}}

\bigskip

\address{
(a) Dipartimento di Fisica G. Occhialini, \\
   Universit\`a di Milano--Bicocca  and INFN, sezione di Milano$^{ 1,2}$
    \\
(b)  Dipartimento di Fisica,  Universit\`a di Milano \\ 
     and INFN, sezione di Milano$^{ 1,2}$
}
\footnotetext{
$^1$mail address: Dipartimento di Fisica, Via Celoria 16, 20133 Milano,
ITALIA.}
\footnotetext{
$^2$e-mail: claudio.destri@@mi.infn.it, emanuele.manfredini@@mi.infn.it
}
\date{November 1999}
\maketitle
\begin{abstract}
The $\lambda \phi^4$ model in a finite volume is studied in the
infinite $N$ limit both at equilibrium and out of equilibrium, with
particular attention to certain fundamental features of the broken
symmetry phase. The numerical solution of the dynamical evolution
equations shows that the zero--mode quantum fluctuations cannot grow
macroscopically large starting from microscopic initial
conditions. Thus we conclude that there is no evidence for a dynamical
Bose--Einstein condensation, in the usual sense. On the other hand,
out of equilibrium the long--wavelength fluctuations do scale with the
linear size of the system, signalling dynamical infrared properties
quite different from the equilibrium ones characteristic of the same
approximation scheme.
\end{abstract}

\newpage

\section{Introduction}\label{int}
In the last few years a great deal of attention has been paid to the
study  of interacting quantum fields out of equilibrium. There are,
in fact, many interesting physical situations in which the
standard S--matrix approach cannot give sensible information about
the behavior of the system, because it evolves through a series of 
highly excited states (i.e., states of finite energy density). 

As an example consider any model of cosmological inflation:
it is not possible to extract precise predictions on physical
observables without including in the treatment the quantum
back--reaction of the field on the space--time geometry and on itself
\cite{devega7,kls,riotto}.

On the side of particle physics, the ultra-relativistic heavy-ion
collisions, scheduled in the forthcoming years at CERN--SPS, BNL--RHIC
and CERN--LHC, are supposed to produce hadron matter at very high
densities and temperatures; in such a regime the usual approach based
on particle scattering cannot be considered a good interpretative tool
at all. To extract sensible information from the theory new
computational schemes are necessary, that go beyond the simple Feynman
diagram expansion. The use of resummation schemes, like the
Hartree--Fock (HF) \cite{tdHF,cactus} approximation and the large $N$
limit (LN) \cite{largen_exp}, or the Hard Thermal Loop resummation for
systems at finite temperature (HTL) \cite{HTL}, can be considered a
first step in this direction. They, in fact, enforce a sum over an
infinite subset of Feynman diagrams that are dominant in a given
region of the parameter space, where the simple truncation of the
usual perturbative series at finite order cannot give sensible
answers.

Quite recently HF and LN have been used in order to clarify some
dynamical aspects of the large $N$ $\phi^4$ theory. For the reader's
benefit and to better motivate this work, we give a very short summary
of the conclusions reached in previous works: i) the early time
evolution is dominated by a so--called ``linear regime'', during which
the energy initially stored in one (or few) modes of the field is
transferred to other modes via either parametric or spinodal
unstabilities, resulting in a large particle production and a
consequent dissipation for the initial condensate \cite{devega2}; ii)
the linear regime stops at a time scale $t_1 \propto \log(\lambda
^{-1})$ (where $\l$ is the quartic coupling constant), by which the
effects of the quantum fluctuation become of the same order as the
classical contribution and the dynamics turns completely non linear
and non perturbative \cite{devega2,relax}; iii) after the time $t_1$
the relaxation occurs via power laws with anomalous dynamical exponent
\cite{relax}; iv) the asymptotic particle distribution, obtained as
the result of the copious particle production at the expenses of the
``classical'' energy, is strongly non-thermal \cite{devega2,relax};
and finally, v) at very large time scale, $t \sim \sqrt{V}$ (where $V$
stands for the volume of the system), the non--perturbative and
non--linear evolution might eventually produce the onset of a novel
form of non--equilibrium Bose--Einstein condensation of the
long--wavelength Goldstone bosons usually present in the broken
symmetry phase of the model \cite{relax,bdvhs}. Another very
interesting result in \cite{bdvhs} concerns the dynamical Maxwell
construction, which reproduces the flat region of the effective
potential in case of broken symmetry as asymptotic fixed points of the
background evolution. Moreover, the LN approximation scheme has been
used to follow the evolution of a initial state characterized by an
occupied spherical shell in momentum space (a spherical 'tsunami'
\cite{tsunami}), around a particular momentum $\left| \vec{k} _0
\right|$, with the following results: i) in a theory where the
symmetry is spontaneously broken at zero density, if we start with a
finite density initial state with restored symmetry, the spinodal
instabilities lead to a dynamical symmetry breaking; ii) the evolution
produces a re-arrangement of the particle distribution towards low
momenta, signalling the onset of Bose condensation; iii) the equation
of state of the asymptotic gas is ultra-relativistic (even if the
distribution is not thermal) \cite{tsunami}.

In this article we present a detailed study, {\em in finite volume},
of dynamical evolution out of equilibrium for the $\Phi^4$ scalar
field in the large $N$ limit. More precisely, we determine how such
dynamics scales with the size of the periodic box containing the
system in the case of uniform backgrounds. This is necessary to
address questions like out--of--equilibrium symmetry breaking and
dynamical Bose--Einstein condensation.

In section \ref{cft} we define the model in finite volume, giving all
the relevant notations and definitions. We also stress the
convexity of the effective potential as an exact result, valid for the
full renormalized theory in any volume.

In section \ref{ln} we derive the large $N$ approximation of the
$O(N)-$invariant version of $\lambda (\bds\phi ^2)^2$ model, according
to the general rules of ref. \cite{yaffe}. In this derivation it
appears evident the essential property of the $N\to\infty$ limit of
being a particular type of {\em classical} limit, so that it leads to
a classical phase space, a classical hamiltonian with associated
Hamilton's equations of motion [see eqs. (\ref{sk2}), (\ref{Nunbgap})
and (\ref{Nbgap})]. We then minimize the hamiltonian function(al) and
determine the conditions for massless Goldstone bosons
(i.e. transverse fluctuations of the field) to form a Bose--Einstein
condensate, delocalizing the vacuum field expectation value (see also
ref. \cite{chkm}). This necessarily requires that the width of the
zero--mode fluctuations becomes macroscopically large, that is of the
order of the volume. Only when the background takes one of the
extremal values proper of symmetry breaking the width of the
zero--mode fluctuations is of order $L^{1/2}$, as typical of a free
massless spectrum.

The study of the lowest energy states of the model is needed for
comparison with the results of the numerical simulations, which show
that the zero--mode width $\s_0$ stays microscopic (that is such that
$\s_0/$volume$\to 0$ when the volume diverges) whenever it starts from
initial conditions in which it is microscopic. Our results, in fact,
show clearly the presence of a time scale $\tau_L$, proportional to
the linear size $L$ of the system, at which finite volume effects
start to manifest. We shall give a very simple physical interpretation
of this time scale in section \ref{numerical}. The important point is that
after $\tau_L$ the zero mode amplitude starts decreasing, then enters
an erratic evolution, but never grows macroscopically large. This
result is at odd with the interpretation of the linear late--time
growth of the zero--mode width as a full dynamical Bose--Einstein
condensation of Goldstone bosons, but is compatible with the ``novel''
form of BEC reported in \cite{relax,bdvhs,tsunami}.

In fact we do find that the size of the low--lying widths at time
$\tau_L$ is of order $L$, to be compared to the equilibrium situation
where they would be of order $L^0$ in the massive case or of order
$L^{1/2}$ in the massless case. Perhaps the denomination
``microscopic'' should be reserved to this two
possibilities. Therefore, since our initial condition are indeed
microscopic in this restricted sense, we do observe in the
out--of--equilibrium evolution a rapid transition to a different
regime intermediate between the microscopic one and the macroscopic
one characteristic of Bose--Einstein condensation. As we shall discuss
more in detail later on, this fully agrees with the result found in
\cite{bdvhs}, that the time--dependent field correlations vanish at
large separations more slowly than for equilibrium free massless
fields (as $r^{-1}$ rather than $r^{-2}$), but definitely faster than
the equilibrium broken symmetry phase characterized by constant
correlations at large distances.

At any rate, when one considers microscopic initial conditions for the
choice of bare mass which corresponds to broken symmetry, the role
itself of symmetry breaking is not very clear in the large $N$
description of the out--of--equilibrium dynamics, making equally
obscure the issues concerning the so--called quantum phase ordering
\cite{bdvhs}. This is because the limit $N\to\infty$ is completely
saturated by gaussian states, which might signal the onset of symmetry
breaking only developing macroscopically large fluctuations. Since
such fluctuations do not appear to be there, the meaning itself of
symmetry breaking, as something happening as times goes on and
accompanied by some kind of phase ordering, is quite unclear. We
postpone to a companion work \cite{finvHF} the discussion about the
possibility of using more comprehensive approximation schemes, that
include some non--gaussian features of the complete theory. As far as
the large $N$ approximation is concerned, we underline that an
important limitation of our approach, as well as of those of the
references mentioned above, is in any case the assumption of a uniform
background. Nonetheless, phenomena like the asymptotic vanishing of
the effective mass and the dynamical Maxwell construction, taking
place in this contest of a uniform background and large $N$ expansion,
are certainly very significant manifestations of symmetry breaking and
in particular of the Goldstone theorem which applies when a continuous
symmetry is broken.

Finally, in section \ref{conclusion} we summarize the results
presented in this article and we sketch some interesting open problems
that we plan to study in forthcoming works.

\section{Cutoff field theory}\label{cft}

We consider the $N-$component scalar field operator $\bds\phi$ in a
$D-$dimensional periodic box of size $L$ and write its Fourier
expansion as customary
\begin{equation*}
	\bds\phi(x)=L^{-D/2}\sum_k \bds\phi_k\,e^{ik\cdot x} 
	\;,\quad \bds\phi_k^\dag = \bds\phi_{-k}
\end{equation*}
with the wavevectors $k$ naturally quantized: $k=(2\pi/L)n$, $n\in\ZZ^D$.
The canonically conjugated momentum $\bds\pi$ has a similar expansion
\begin{equation*}
	\bds\pi(x)=L^{-D/2}\sum_k \bds\pi_k\,e^{ik\cdot x} 
	\;,\quad \bds\pi_k^\dag = \bds\pi_{-k}
\end{equation*}
with the commutation rules $[\phi_k^\a\,,\pi_{-k'}^\b]=
i\,\delta^{(D)}_{kk'}\,\delta^{\a\b}$.  The introduction of a finite
volume should be regarded as a regularization of the infrared
properties of the model, which allows to ``count'' the different field
modes and is needed especially in the case of broken symmetry. In
fact, all the results we have summarized in section \ref{int}, have
been obtained simulating the system directly in infinite volume, where
the evolution equations contain momentum integrals, that must be
computed numerically by a proper, but nonetheless rather arbitrary,
discretization in momentum space. Of course, the final result should
be as insensitive as possible to the particular choice of the
integration grid. In such a situation, the definition of a ``zero''
mode and the interpretation of its late time behavior might not be
rigorous enough, unless, for some reason, it turns out that a
particular mode requires a different treatment compared to the
others. In order to understand this point, it is necessary to put the
system in a finite volume (a box of size $V$); the periodic boundary
conditions let us single out the zero mode in a rigorous way and thus
we can carefully analyze its scaling properties with $V$ and get some
information on the infinite volume limit.

To regularize also the ultraviolet behavior, we restrict the sums
over wavevectors to the points lying within the $D-$dimensional sphere
of radius $\Lambda$, that is $k^2\le\Lambda^2$, with $\N=\Lambda
L/2\pi$ some large integer. Clearly we have reduced the original
field--theoretical problem to a quantum--mechanical framework with
finitely many (of order $\N^{D-1}$) degrees of freedom.

The $\phi^4$ Hamiltonian reads
\begin{equation}\label{lN_ham}
\begin{split}
  H &=\frac12\int d^Dx \left[\bds\pi^2 + (\partial\bds\phi)^2 +
  \mbare\,\bds\phi^2 + \lbare( \bds\phi^2)^2 \right] =
  \frac12\sum_k\left[\bds\pi_k \!\cdot \bds\pi_{-k} +(k^2+
  \mbare)\,\bds\phi_k \!\cdot \bds\phi_{-k}\right] +\\
  &+\dfrac\lbare{4L^D} \sum_{k_1,k_2,k_3,k_4} (\bds\phi_{k_1}\!\!\cdot
  \bds\phi_{k_2})(\bds\phi_{k_3} \!\!\cdot \bds\phi_{k_4})
  \,\delta^{(D)}_{k_1+k_2+k_3+k_4,0}
\end{split}
\end{equation}
where $\mbare$ and $\lbare$ should depend on the UV cutoff $\Lambda$
in such a way to guarantee a finite limit $\Lambda\to\infty$ for all
observable quantities. As is known
\cite{devega2,wilson%,froh,latt,broken
}, this implies triviality
(that is vanishing of renormalized vertex functions with more than two
external lines) for $D>3$ and very likely also for $D=3$. In the
latter case triviality is manifest in the one--loop approximation and
in large$-N$ limit due to the Landau pole.  For this reason we shall
keep $\Lambda$ finite and regard the $\phi^4$ model as an effective
low--energy theory (here low--energy means practically all energies
below Planck's scale, due to the large value of the Landau pole for
renormalized coupling constants of order one or less).

We shall work in the wavefunction representation where
$\braket{\bds\varphi}\Psi=\Psi(\bds\varphi)$ and
\begin{equation*}
	(\bds\phi_0\Psi)(\bds\varphi)=\bds\varphi_0 \Psi(\bds\varphi)
	  \;,\quad 
	(\bds\pi_0\Psi)(\bds\varphi)= -i\pdif{}{\bds\varphi_0}\Psi(\bds\varphi)
\end{equation*}
while for $k>0$ (in lexicographic sense)
\begin{equation*}
	(\bds\phi_{\pm k}\Psi)(\bds\varphi)=\dfrac1{\sqrt2}\left(
	\bds\varphi_k\pm i\,\bds\varphi_{-k}\right)\Psi(\bds\varphi)
	  \;,\quad 
	(\bds\pi_{\pm k}\Psi)(\bds\varphi)= \dfrac1{\sqrt2} \left(-i
	\pdif{}{\bds\varphi_k}\pm \pdif{}{\bds\varphi_{-k}}\right)\Psi(\bds\varphi)
\end{equation*}
Notice that by construction the variables $\bds\varphi_k$ are all real.
Of course, when either one of the cutoffs are removed, the wave function
$\Psi(\bds\varphi)$ acquires infinitely many arguments and becomes what is
usually called a {\em wavefunctional}.

In practice, the problem of studying the dynamics of the $\phi^4$
field out of equilibrium consists now in trying to solve the
time-dependent Schroedinger equation given an initial wavefunction
$\Psi(\bds\varphi,t=0)$ that describes a state of the field far away from
the vacuum. By this we mean a non--stationary state that, in the
infinite volume limit $L\to\infty$, would lay outside the particle
Fock space constructed upon the vacuum. This approach could be
generalized in a straightforward way to mixtures described by density
matrices, as done, for instance, in \cite{hkmp,bdv,devega3}. Here we shall
restrict to pure states, for sake of simplicity and because all
relevant aspects of the problem are already present in this case.

It is by now well known \cite{devega2} that perturbation theory is not
suitable for the purpose stated above. Due to parametric resonances
and/or spinodal instabilities there are modes of the field that grow
exponentially in time until they produce non--perturbative effects for
any coupling constant, no matter how small. On the other hand, only
few, by now standard, approximate non--perturbative schemes are
available for the $\phi^4$ theory, and to these we have to resort
after all.  We shall consider here only the large $N$ expansion to
leading order, remanding to another work the definition of a
time-dependent Hartree--Fock (tdHF) approach \cite{finvHF} (a
generalization of the treatment given, for instance, in
\cite{tdHF}). In fact these two methods are very closely related, as
shown in \cite{inomogeneo}, where several techniques to derive
reasonable dynamical evolution equations for non--equilibrium $\phi^4$
are compared. 

We close this section by stressing that the introduction of both a UV
and IR cutoff allows to easily derive the well--known rigorous result
concerning the flatness of the effective potential.  In fact
$V_{\text{eff}}(\bar\phi)$ is a convex analytic function in a finite
neighborhood of $\bar\phi=0$, as long as the cutoffs are present, due
to the uniqueness of the ground state. In the infrared limit
$L\to\infty$, however, $V_{\text{eff}}(\bar\phi)$ might flatten around
$\bar\phi=0$. Of course this possibility would apply in case of
spontaneous symmetry breaking, that is for a double--well classical
potential. This is a subtle and important point that will play a
crucial role later on, even if the effective potential is relevant for
the static properties of the model rather than the dynamical evolution
out of equilibrium that interests us here. In fact, the dynamical
evolution in QFT is governed by the CTP effective action
\cite{schwinger,zshy} and one might expect that, although non--local
in time, it asymptotically reduces to a multiple of the effective
potential for trajectories of $\bar\phi(t)$ with a fixed point at
infinite time. In such case there should exist a one--to--one
correspondence between fixed points and minima of the effective
potential.

\section{Large $N$ expansion at leading order}\label{ln}
\subsection{Definitions}
In this section we consider the standard non--perturbative approach to
the $\phi^4$ model which is applicable also out of equilibrium, namely
the large $N$ method as presented in \cite{chkmp}. However we shall follow
a different derivation which makes the gaussian nature of the 
$N\to\infty$ limit more explicit.

It is known that the theory described by the Hamiltonian (\ref{lN_ham})
is well behaved for large $N$, provided that the quartic coupling
constant $\lbare$ is rescaled with $1/N$. For example, it is possible
to define a perturbation theory, based on the small expansion
parameter $1/N$, in the framework of which one can compute any
quantity at any chosen order in $1/N$. From the diagrammatic point of
view, this procedure corresponds to a resummation of the usual
perturbative series that automatically collects all the graphs of a
given order in $1/N$ together \cite{largen_exp}. Moreover, it has been
established since the early 80's that the leading order approximation
(that is the strict limit $N\to\infty$) is actually a classical limit
\cite{yaffe}, in the sense that there exists a classical system (i.e.,
a classical phase space, a Poisson bracket and a classical
Hamiltonian) whose dynamics controls the evolution of all fundamental
quantum observables, such as field correlation functions, in the
$N\to\infty$ limit. For instance, from the absolute minimum of the
classical Hamiltonian one reads the energy of the ground state, while
the spectrum is given by the frequencies of small oscillations about
this minimum, etc. etc.. We are here interested in finding an
efficient and rapid way to compute the quantum evolution equations for
some observables in the $N \to \infty$ limit, and we will see that
this task is easily accomplished just by deriving the canonical
Hamilton equations from the large $N$ classical Hamiltonian.

Following Yaffe \cite{yaffe}, we write the quantum mechanical
hamiltonian as 
\begin{equation}\label{hAC}
	H = N h (A\,,C) 
\end{equation}
in terms of the square matrices $A$, $C$ with operator entries
({\bf $\bds\varpi_k$} is the canonical momentum conjugated to the real mode
{\bf $\bds\varphi_k$})
\begin{equation*}
	A_{kk'} = \frac1{N} \bds\varphi_k \!\cdot \bds\varphi_{k'}
   \;,\quad C _{kk'} = \frac1{N} \bds\varpi_k\!\cdot \bds\varpi_{k'}
\end{equation*}
These are example of ``classical'' operators, whose two-point correlation
functions factorize in the $N\to\infty$ limit. This can be shown by
considering the coherent states 
\begin{equation}
	\Psi_{z,q,p}( \bds\varphi) = C(z) \exp\left[ 
	i\sum_k \bds p_k\cdot \bds\varphi_k - \frac1{2N} \sum_{kk'} z_{kk'} 
	(\bds\varphi_k-\bds q_k) \cdot(\bds\varphi_{k'}-\bds q_{k'}) \right]
\label{coh_states}
\end{equation}
where the complex symmetric matrix $z$ has a positive definite real
part while $\bds p_k$ and $\bds q_k$ are real and coincide,
respectively, with the coherent state expectation values of $\bds
\varpi_k$ and $\bds \varphi_k$. As these parameters take  all their
possible values, the coherent states form an overcomplete set in the
cutoff Hilbert space of the model. The crucial property which ensures
factorization is that they become all orthogonal in the $N\to\infty$
limit. Moreover one can show \cite{yaffe} that the 
coherent states parameters form a classical phase space with Poisson
brackets 
\begin{equation*}
	\left\{q_k^i\,,p_{k'}^j\right\}_{\rm P.B.} = \delta_{kk'}
	\delta^{ij} \;,\quad
	\left\{w_{kk'}\,,v_{qq'}\right\}_{\rm P.B.} = 
	\delta _{kq} \delta _{k'q'} + \delta _{kq'} \delta _{k'q}
\end{equation*}
where $w$ and $v$ reparametrize $z$ as $z=\frac12
w^{-1}+i\,v$. It is understood that the dimensionality of the vectors
$\bds q_k$ and $\bds p_k$ is arbitrary but finite [that is, only a finite
number, say $n$, of pairs $(\varphi_k^i\,,\varpi_k^i)$ may take a nonvanishing
expectation value as $N\to\infty$].

Once applied to the classical operators $A_{kk'}$ and
$C_{kk'}$ the large $N$ factorization allow to obtain the classical
hamiltonian by simply replacing $A$ and $C$ in eq. (\ref{hAC}) by the
coherent expectation values
\begin{equation*}
	\vev{A_{kk'}} = \bds q_k\cdot \bds q_{k'} + w_{kk'} \;,\quad
	\vev{C_{kk'}} = \bds p_k\cdot \bds p_{k'} + 
	(v\,w\,v)_{kk'} + \frac14 ( w^{-1})_{kk'}
\end{equation*}
In our situation, having assumed a uniform background expectation
value for $\bds \phi$, we have $\bds q_k=\bds p_k=0$ for all
$k\neq 0$; moreover, translation invariance implies that 
$w$ and $v$ are diagonal matrices, so that we may set
\begin{equation*}
	w_{kk'} = \s_k^2\,\delta_{kk'} \;,\quad 
	v_{kk'}=\frac{s_k}{\s_k}\,\delta_{kk'}
\end{equation*}
in term of the canonical couples $(\s_k,s_k)$ which satisfy 
$\{\s_k\,,s_{k'}\}_{\rm P.B.}=\delta_{kk'}$. Notice that the  $\s_k$ are
just the widths (rescaled by $N^{-1/2}$) of the $O(N)$ symmetric
and translation invariant gaussian coherent states.

Thus we find the classical hamiltonian
\begin{equation*}
	h_{\rm cl} =\frac12(\bds p_0^2 +\mbare\, \bds q^2_0) +
	\frac12\sum_k\left[ s_k^2 +(k^2+\mbare)\s_k^2 + \frac1{4\s^2_k}
	\right] + \frac\lbare{4L^D} \left(\bds q_0^2+\sum_k\s_k^2\right)^2
\end{equation*}
where by Hamilton's equations of motion $\bds p_0=\dot{\bds q}_0$ and
$s_k=\dot\s_k$. The corresponding conserved energy density
$\E=L^{-D}h_{\rm cl}$ may be written
\begin{equation}\label{ElargeN}
\begin{split}
	\E &= \T +\V \;,\quad \T = \frac12\dot{\bar{\bds\phi}}^2 +
	\frac1{2L^D}\sum_k \dot\s_k^2  \\ \V &= \frac1{2L^D}\sum_k\left(
	 k^2\,\s_k^2 + \frac1{4\s^2_k}\right) +V(\bar{\bds\phi}^2
	+\Sigma) \;,\quad \Sigma=\frac1{L^D}\sum_k\s_k^2
\end{split}
\end{equation}
where $\bar{\bds\phi}=L^{-D/2}\bds q_0$ and $V$ is
the $O(N)-$invariant quartic potential regarded as a function of
$\bds\phi^2$, that is $V(z)=\frac12\mbare z + \frac14\lbare z^2$. It
is worth noticing that eq. (\ref{ElargeN}) would apply as is to
generic $V(z)$.

\subsection{Static properties}
Let us consider first the static aspects embodied in the effective
potential $V_{\text{eff}}(\bar{\bds\phi})$, that is the minimum of the
potential energy $\V$ at fixed $\bar{\bds\phi}$.  We first define in a
precise way the unbroken symmetry phase, in this large $N$ context, as
the case when $V_{\text{eff}}(\bar{\bds\phi})$ has a unique minimum at
$\bar{\bds\phi}=0$ in the limit of infinite volume. Minimizing $\V$
w.r.t. $\s_k$ yields
\begin{equation}\label{massive}
\begin{split}
	\s^2_k=\frac1{2\sqrt{k^2+M^2}} \;,\quad
	M^2 &= \mbare + 2\,V'(\bar{\bds\phi}^2+\Sigma) \\
	&= \mbare + \lbare\bar{\bds\phi}^2+ \frac\lbare{L^D}
	\sum_k\frac1{2\sqrt{k^2+M^2}}
\end{split}
\end{equation}
that is the widths characteristic of a free theory with
self--consistent mass $M$ fixed by the gap equation. By the assumption
of unbroken symmetry, when $\bar{\bds\phi}=0$ and at infinite volume
$M$ coincides with the equilibrium mass $m$ of the theory, that may be
regarded as independent scale parameter. Since in the limit
$L\to\infty$ sums are replaced by integrals
\begin{equation*}
	\Sigma \to \int_{k^2\le\Lambda^2} \dfrac{d^Dk}{(2\pi)^D} \s_k^2
\end{equation*}
we obtain the standard bare mass parameterization
\begin{equation}\label{Nm2ren}
	\mbare= m^2 - \lbare I_D(m^2,\Lambda) \;,\quad 
	I_D(z,\Lambda) \equiv \int_{k^2\le\Lambda^2} 
	\dfrac{d^Dk}{(2\pi)^D} \dfrac1{2\sqrt{k^2+z}}
\end{equation}
and the renormalized gap equation
\begin{equation}\label{rengapN}
	M^2 = m^2+ \l\,\bar\phi^2+\l \left[ I_D(M^2,\Lambda)-
	I_D(m^2,\Lambda) \right]_{\text{finite}} 
\end{equation}
which implies, when $D=3$,
\begin{equation}\label{lrenN}
	\lbare =\l \left(1-\frac{\l}{8\pi^2}
	    \log\frac{2\Lambda}{m\sqrt{e}} \right)^{\!\!-1} 
\end{equation}
with a suitable choice of the finite part. No coupling constant
renormalization occurs instead when $D=1$. The renormalized gap
equation (\ref{rengapN}) may also be written quite concisely
\begin{equation}\label{nice}
	\frac{M^2}{\hat\l(M)} = \frac{m^2}{\hat\l(m)} + \,\bar{\bds\phi}^2
\end{equation}
in terms of the one--loop running couplings constant
\begin{equation*}
	\hat\l(\mu) = \l \left[ 1 - \frac{\l}{8\pi^2} 
	\log\frac{\mu}m \right]^{-1} \;,\quad \hat\l(m)=\l
	\;,\quad \hat\l(2\Lambda\,e^{-1/2}) =\lbare
\end{equation*}
It is the Landau pole in $\hat\l(2\Lambda\,e^{-1/2})$ that actually
forbids the limit $\Lambda\to\infty$. Hence we must keep the cutoff
finite and smaller than $\Lambda_{\text{pole}}$, so that the theory
does retain a slight inverse--power dependence on it. At any rate,
there exists a very wide window where this dependence is indeed very
weak for couplings of order one or less, since
$\Lambda_{\text{pole}}=\frac12 m\exp(1/2+8\pi^2/\l)\gg m$.  Moreover,
we see from eq. (\ref{nice}) that for $\sqrt\l|\bar\phi|$ much smaller
than the Landau pole there are two solutions for $M$, one
``physical'', always larger than $m$ and of the same order of
$m+\sqrt\l|\bar\phi|$, and one ``unphysical'', close to the Landau
pole.

One can now easily verify that the effective potential has indeed a
unique minimum in $\bar{\bds\phi}=0$, as required. In fact, if we
assign arbitrary $\bar\phi-$dependent values to the widths $\s_k$,
(minus) the effective force reads
\begin{equation}\label{Nefforce}
	\der{}{\bar\phi^i}\V(\bar{\bds\phi},\{\s_k(\bar{\bds\phi})\})
	= M^2\,\bar\phi^i + \sum_k \pdif\V{\s_k}
	\der{\s_k}{\bar\phi^i}
\end{equation} 
and reduces to $M^2\,\bar\phi^i$ when the widths are extremal as
in eq. (\ref{massive}); but $M^2$ is positive for unbroken symmetry
and so $\bar{\bds\phi}=0$ is the unique minimum. 

We define the symmetry as broken whenever the infinite volume
$V_{\text{eff}}$ has more than one minimum.  Of course, as long as $L$
is finite, $V_{\text{eff}}$ has a unique minimum in
$\bar{\bds\phi}=0$, because of the uniqueness of the ground state in
Quantum Mechanics, as already discussed in section \ref{cft}. Let us
therefore proceed more formally and take the limit $L\to\infty$
directly on the potential energy $\V$. It reads
\begin{equation*}
	\V = \frac12 \int_{k^2\le\Lambda^2} \dfrac{d^Dk}{(2\pi)^D}
	 \left(k^2\,\s_k^2 + \frac1{4\s^2_k}\right) 
	+V(\bar{\bds\phi}^2 +\Sigma) \;,\quad  
	\Sigma =\int_{k^2\le\Lambda^2} \dfrac{d^Dk}{(2\pi)^D}\,\s_k^2
\end{equation*}
where we write for convenience the tree--level potential $V$ in the
positive definite form $V(z)=\frac14\lbare(z+\mbare/\lbare)^2$.
$\V$ is now the sum of two positive definite terms.
Suppose there exists a configuration such that 
$V(\bar{\bds\phi}^2 +\Sigma)=0$ and the first term in $\V$ is at
its minimum. Then this is certainly the absolute minimum of $\V$.
This configuration indeed exists at infinite volume when $D=3$:
\begin{equation}\label{minimum}
	\s^2_k=\frac1{2|k|} \;,\quad  \bar{\bds\phi}^2 =v^2
	\;,\quad \mbare = -\lbare\left[ v^2+ I_3(0,\Lambda) \right]
\end{equation}
where the nonnegative $v$ should be regarded as an independent
parameter fixing the scale of the symmetry breaking. It replaces the
mass parameter $m$ of the unbroken symmetry case: now the theory is
massless in accordance with Goldstone theorem.  On the contrary, if
$D=1$ this configuration is not allowed due to the infrared
divergences caused by the massless nature of the width spectrum. This
is just the standard manifestation of Mermin--Wagner--Coleman theorem
that forbids continuous symmetry breaking in a two--dimensional
space--time \cite{mwc}.

At finite volumes we cannot minimize the first term in $\V$ since this
requires $\s_0$ to diverge, making it impossible to keep
$V(\bar{\bds\phi}^2 +\Sigma)=0$. In fact we know that the
uniqueness of the ground state with finitely many degrees of freedom
implies the minimization equations (\ref{massive}) to hold always
true with a $M^2$ strictly positive.  Therefore, broken symmetry
should manifest itself as the situation in which the equilibrium value
of $M^2$ is a positive definite function of $L$ which vanishes in the
$L\to\infty$ limit.

We can confirm this qualitative conclusion as follows. We assume that
the bare mass has the form given in eq. (\ref{minimum}) and that 
$\bar{\bds\phi}^2 =v^2$ too. Minimizing the potential energy
leads always to the massive spectrum, eq. (\ref{massive}), with the
gap equation
\begin{equation}\label{gapagain}
	\frac{M^2}\lbare = \frac1{2L^3M} + \frac1{2L^3}
	\sum_{k\neq0}\frac1{\sqrt{k^2+M^2}} - \frac{\Lambda^2}{8\pi^2}
\end{equation}
If $M^2>0$ does not vanish too fast for large volumes, or stays even
finite, then the sum on the modes has a behavior similar to the
corresponding infinite volume integral: there is a quadratic
divergence that cancels the infinite volume contribution, and a
logarithmic one that renormalizes the bare coupling. The direct
computation of the integral would produce a term containing the
$M^2\log(\Lambda/M)$. This can be split into
$M^2[\log(\Lambda/v)-\log(M/v)]$ by using $v$ as mass scale. The first
term renormalizes the coupling correctly, while the second one
vanishes if $M^2$ vanishes in the infinite volume limit.

When $L\to\infty$, the asymptotic solution of (\ref{gapagain}) reads
\begin{equation*}
	M = \left( \frac{\l}{2}\right) ^{1/3} L ^{-1} + \rm{h.o.t.}
\end{equation*}
that indeed vanishes in the limit. Note also that the exponent is
consistent with the assumption made above that $M$ vanishes slowly
enough to approximate the sum over $k\neq 0$ with an integral with the
same $M$.

Let us now consider a state whose field expectation value
$\bar{\bds\phi}^2$ is different from $v^2$. If $\bar{\bds\phi}^2 > v$,
the minimization equations (\ref{massive}) leads to a positive squared
mass spectrum for the fluctuations, with $M^2$ given
self--consistently by the gap equation. On the contrary, as soon as
$\bar{\bds\phi}^2 < v^2$, one immediately see that a positive $M^2$
cannot solve the gap equation
\begin{equation*}
	M^2 = \lbare \left( \bar{\bds\phi}^2 -v^2 + \frac{\s_0^2}{L^3} +
	\frac1{2L^3} \sum_{k\neq0}\frac1{\sqrt{k^2+M^2}} - 
	\frac{\Lambda^2}{8 \pi ^2} \right)
\end{equation*}
if we insist on the requirement that $\s_0$ not be macroscopic. In fact,
the r.h.s. of the previous equation is negative, no matter which
positive value for the effective mass we choose, at least for $L$
large enough. But nothing prevent us to consider a static
configuration for which the amplitude of the zero mode is
macroscopically large (i.e. it rescales with the volume $L^3$).
Actually, if we choose
\begin{equation*}
	\frac{\s _0 ^2}{L^3} =  v ^2 - \bar{\bds\phi}^2 + \frac1{2L^3M}
\end{equation*}
we obtain the same equation as we did before and the same value for
the potential, that is the minimum, in the limit $L \to \infty$. Note
that at this level the effective mass $M$ needs not to have the same
behavior in the $L \to \infty$ limit, but it is free of rescaling
with a different power of $L$. We can be even more precise: we isolate
the part of the potential that refers to the zero mode width $\s_0$
($\Sigma'$ does not contain the $\s_0$ contribution)
\begin{equation*}
\frac12 \left[ \mbare + \lbare \left( \bar{\bds\phi}^2 +\Sigma' \right)
\right] \frac{\s _0 ^2}{L ^3} + \frac{\lbare}4 \frac{\s _0 ^4}{L ^6} +
\frac{1}{8 L ^3 \sigma _0 ^2}
\end{equation*}
and we minimize it, keeping $\bar{\bds\phi}^2$ fixed. The minimum is
attained at $t=\s_0^2/L^3$ solution of the cubic equation
\begin{equation*}
\lbare t^3 + \a \lbare t^2 - \tfrac14 L^{-6} = 0
\end{equation*}
where $\a = \bar{\bds\phi}^2 - v ^2 + \Sigma ' - I _3 \left( 0 ,
\Lambda \right)$. Note that $\lbare \alpha$ depends on $L$ and it has
a finite limit in infinite volume: $\l (\bar{\bds\phi}^2 - v ^2)$. The
solution of the cubic equation is
\begin{equation*}
	\lbare t = \lbare ( v ^2 - \bar{\bds\phi}^2 ) + 
	\tfrac14 [L^3(v^2 -\bar{\bds\phi}^2)]^{-2} + \rm{h.o.t.}
\end{equation*}
from which the effective mass can be identified as proportional
to $L^{-3}$. The stability equations for all the other modes can now
be solved by a massive spectrum, in a much similar way as before. 

Since $\s_0$ is now macroscopically large, the infinite volume limit
of the $\s_k$ distribution (that gives a measure of the {\em transverse}
fluctuations in the $O(N)$ model) develop a $\delta-$like singularity, 
signalling  a Bose condensation of the Goldstone bosons: 
\begin{equation}\label{BE}
	\s_k^2 = ( v ^2 - \bar{\bds\phi}^2 )\,\delta^{(D)}(k)+
	\frac1{2k}
\end{equation}
At the same time it is evident that the minimal potential energy is
the same as when $\bar{\bds\phi}^2=v^2$, that is  the effective
potential flattens, in accord with the Maxwell construction.

Eq. (\ref{BE}) corresponds in configuration space to the $2-$point
correlation function
\begin{equation}\label{2pN}
	\lim_{N \to \infty} \frac{\vev{\bds{\phi}(x)
	\cdot\bds{\phi}(y)}}{N} = \bar{\bds{\phi}}^2+ \int\frac{d^Dk}
	{(2\pi)^D}\,\s_k^2\, e^{ik\cdot(x-y)}= C(\bar{\bds{\phi}}^2)
	+\Delta_D(x-y)
\end{equation}
where $\Delta_D(x-y)$ is the massless free--field equal--time
correlator, while  
\begin{equation}\label{broken}
	C(\bar{\bds{\phi}}^2)= v^2\,\Theta(v^2-\bar{\bds{\phi}}^2) +
	\bar{\bds{\phi}}^2\, \Theta(\bar{\bds{\phi}}^2-v^2) =
	\text{max}(v^2,\bar{\bds{\phi}}^2)
\end{equation}
This expression can be extended to unbroken symmetry by
setting in that case $C(\bar{\bds{\phi}}^2)=\bar{\bds{\phi}}^2$.

Quite evidently, when eq. (\ref{broken}) holds, symmetry breaking can
be inferred from the limit $|x-y|\to\infty$, if clustering is assumed
\cite{zj,allthat}, since $\Delta_D(x-y)$ vanishes for large
separations. Of course this contradicts the infinite volume limit of
the finite--volume definition, $\bar{\bds\phi}=\lim_{N\to\infty}
N^{-1/2}\vev{\bds\phi(x)}$, except at the extremal points
$\bar{\bds{\phi}}^2=v^2$.

In fact the $L\to\infty$ limit of the finite volume states with
$\bar{\bds\phi}^2<v^2$ do violate clustering, because they are linear
superpositions of vectors belonging to superselected sectors and
therefore they are indistinguishable from statistical mixtures. We can
give the following intuitive picture for large $N$. Consider any one
of the superselected sectors based on a physical vacuum with
$\bar{\bds\phi}^2=v^2$. By condensing a macroscopic number of
transverse Goldstone bosons at zero--momentum, one can build coherent
states with rotated $\bar{\bds\phi}$. By incoherently averaging over
such rotated states one obtains new states with field expectation
values shorter than $v$ by any prefixed amount.  In the large $N$
approximation this averaging is necessarily uniform and is forced upon
us by the residual $O(N-1)$ symmetry.

\subsection{Out--of--equilibrium dynamics}\label{ooedN}

We now turn to the dynamics out of equilibrium in this large $N$
context. It is governed by the equations of motion derived from the
total energy density $\E$ in eq. (\ref{ElargeN}), that is
\begin{equation}\label{sk2}
	\der{^2\bar{\bds\phi}}{t^2} = -M^2\,\bar{\bds\phi}
	\;,\quad \der{^2 \s_k}{t^2}= -(k^2+M^2)\,\s_k + \frac1{4 \s_k^3}
\end{equation}
where the generally time--dependent effective squared mass $M^2$ is
given by
\begin{equation}\label{Nunbgap}
	M^2 = m^2 + \lbare
	\left[\bar{\bds\phi}^2+\Sigma-I_D(m^2,\Lambda)\right]
\end{equation}
in case of unbroken symmetry and
\begin{equation}\label{Nbgap}
	M^2 = \lbare \left[\bar{\bds\phi}^2- v^2 +\Sigma-I_3(0,\Lambda)\right]
\end{equation}
for broken symmetry in $D=3$.

At time zero, the specific choice of initial conditions for $\s_k$ that
give the smallest energy contribution, that is
\begin{equation}\label{inis}
	\dot \s_k=0 \;,\quad \s_k^2=\frac1{2\sqrt{k^2+M^2}}
\end{equation}
turns eq. (\ref{Nunbgap}) into the usual gap equation
(\ref{massive}). For any value of $\bar\phi$ this equation has one
solution smoothly connected to the value $M=m$ at $\bar\phi=0$.  Of
course other initial conditions are possible. The only requirement is
that the corresponding energy must differ from that of the ground
state by an ultraviolet finite amount, as it occurs for the choice
(\ref{inis}). In fact this is guaranteed by the gap equation
itself, as evident from eq. (\ref{Nefforce}): when the widths $\s_k$ 
are extremal the effective force is finite, and therefore so are all
potential energy differences.

This simple argument needs a refinement in two respects. 

Firstly, in case of symmetry breaking the formal energy minimization
w.r.t. $\s_k$ leads always to eqs. (\ref{inis}), but these are
acceptable initial conditions only if the gap equation that follows
from eq. (\ref{Nbgap}) in the $L\to\infty$ limit, namely
\begin{equation}\label{Nbgap2}
	M^2 = \lbare \left[\bar{\bds\phi}^2-v^2 +
	I_D(M^2,\Lambda)-I_D(0,\Lambda) \right] 
\end{equation}
admits a nonnegative, physical solution for $M^2$. 

Secondly, ultraviolet finiteness only requires that the sum over $k$
in eq. (\ref{Nefforce}) be finite and this follows if eq. (\ref{inis})
holds at least for $k$ large enough, solving the issue raised in the
first point: negative $M^2$ are allowed by imposing a new form of gap
equation
\begin{equation}\label{newgap}
	M^2 = \lbare \left[\bar{\bds\phi}^2-v^2 + \frac1{L^D} 
	\sum_{k^2<|M^2|}\s_k^2 + \frac1{L^D}\sum_{k^2>|M^2|}
	\frac1{2\sqrt{k^2-|M^2|}} -I_D(0,\Lambda) \right] 
\end{equation}
where all $\s_k$ with $k^2<|M^2|$ are kept free (but all by hypothesis
microscopic) initial conditions. Of course there is no energy
minimization in this case. To determine when this new form is
required, we observe that, neglecting the inverse--power corrections
in the UV cutoff we may write eq. (\ref{Nbgap2}) in the following
form 
\begin{equation}\label{NiceN}
	\frac{M^2}{\hat\l(M)} = \bar{\bds\phi}^2 -v^2
\end{equation}
There exists a positive solution $M^2$ smoothly connected to the
ground state, $\bar{\bds\phi}^2=v^2$ and $M^2=0$, only provided
$\bar{\bds\phi}^2\ge v^2$. So, in the large $N$ limit, as soon as we start
with $\bar{\bds\phi}^2\le v^2$, we cannot satisfy the gap equation with a
positive value of $M^2$.

Once a definite choice of initial conditions is made, the system of
differential equations (\ref{sk2}), (\ref{Nunbgap}) or (\ref{Nbgap})
can be solved numerically with standard integration algorithms.  This
has been already done by several authors \cite{devega2,relax,bdvhs},
working directly in infinite volume, with the following general
results. In the case of unbroken symmetry it has been established
that the $\s_k$ corresponding to wavevectors $k$ in the so--called
forbidden bands with parametric resonances grow exponentially in time
until their growth is shut off by the back--reaction. For broken
symmetry it is the region in $k-$space with the spinodal instabilities
caused by an initially negative $M^2$, whose widths grow exponentially
before the back--reaction shutoff.  After the shutoff time the
effective mass tends to a positive constant for unbroken symmetry and
to zero for broken symmetry (in D=3), so that the only width with a
chance to keep growing indefinitely is $\s_0$ for broken symmetry.

Of course, in all these approaches the integration over modes in the
back--reaction $\Sigma$ cannot be done exactly and is always replaced
by a discrete sum of a certain type, depending on the details of the
algorithms. Hence there exists always an effective infrared cutoff,
albeit too small to be detectable in the numerical outputs.  A
possible troublesome aspect of this is the proper identification of
the zero--mode width $\s_0$. Even if a (rather arbitrary) choice of
discretization is made where a $\s_0$ appears, it is not really
possible to determine whether during the exponential growth or after
such width becomes of the order of the volume.  Our aim is just to
answer this question and therefore we perform our numerical
evolution in finite volumes of several growing sizes.
Remanding to the appendix for the details of our method, we summarize
our results in the next subsection.

\subsection{Numerical results}\label{numerical}
After a careful study in $D=3$ of the scaling behavior of the
dynamics with respect to different values of $L$, the linear size of
the system, we reached the following conclusion: there exist a
$L-$dependent time, that we denote by $\tau_L$, that splits
the evolution in two parts; for $t \leq \tau_L$, the
behavior of the system does not differ appreciably from its
counterpart at infinite volume, while finite volume effects abruptly
alter the evolution as soon as $t$ exceeds $\tau_L$; moreover
\begin{itemize}
\item $\tau_L$ is proportional to the linear size of the
box $L$ and so it rescales as the cubic root of the volume.
\item $\tau_L$ does not depend on the value of the quartic
coupling constant $\lambda$, at least in a first approximation.
\end{itemize}

\begin{figure}
\epsfig{file=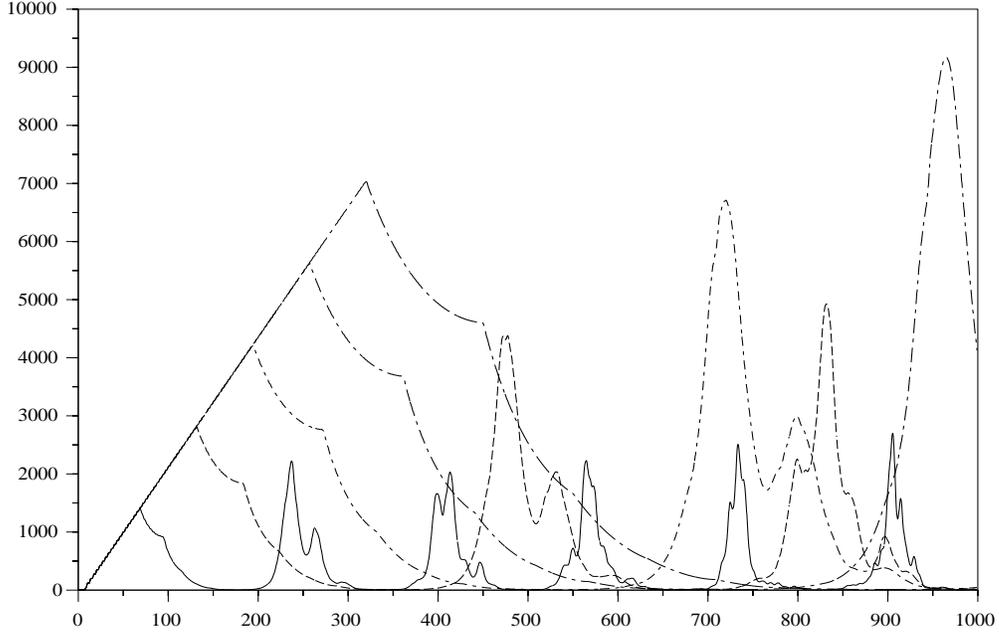,height=10cm,width=15cm}
\caption{\it Zero--mode amplitude evolution for 
different values of the size
$\frac{L}{2\pi}=20,40,60,80,100$, for $\lambda = 0.1$ and broken symmetry,
with $\bar\phi=0$. }\label{fig:m0}
\end{figure}

\begin{figure}
\epsfig{file=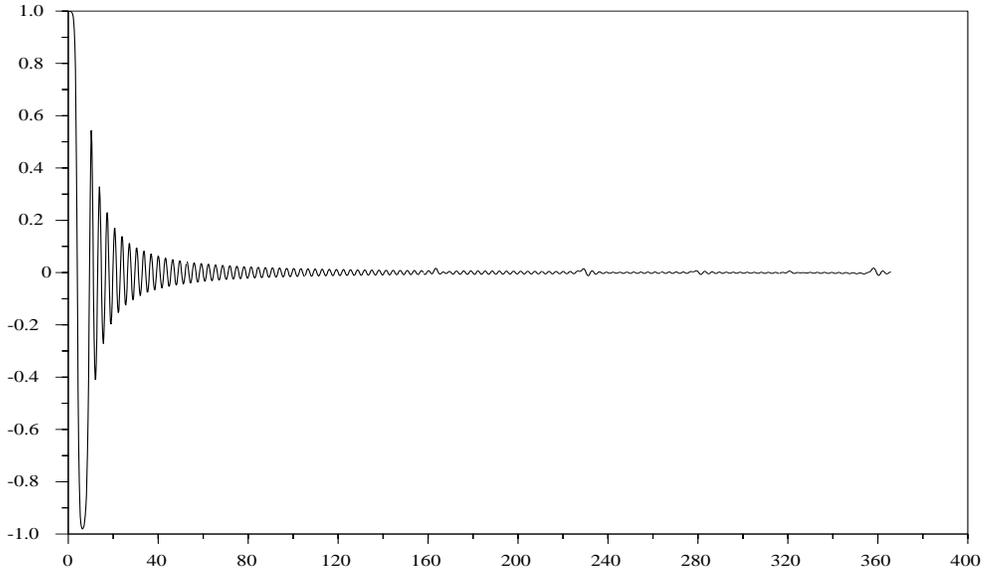,height=9cm,width=15cm}
\caption{\it Time evolution of the squared effective mass $M^2$ in broken
symmetry, for $\frac{L}{2\pi}=50$ and $\lambda=0.1$. }\label{fig:mass2}
\end{figure}

\begin{figure}
\epsfig{file=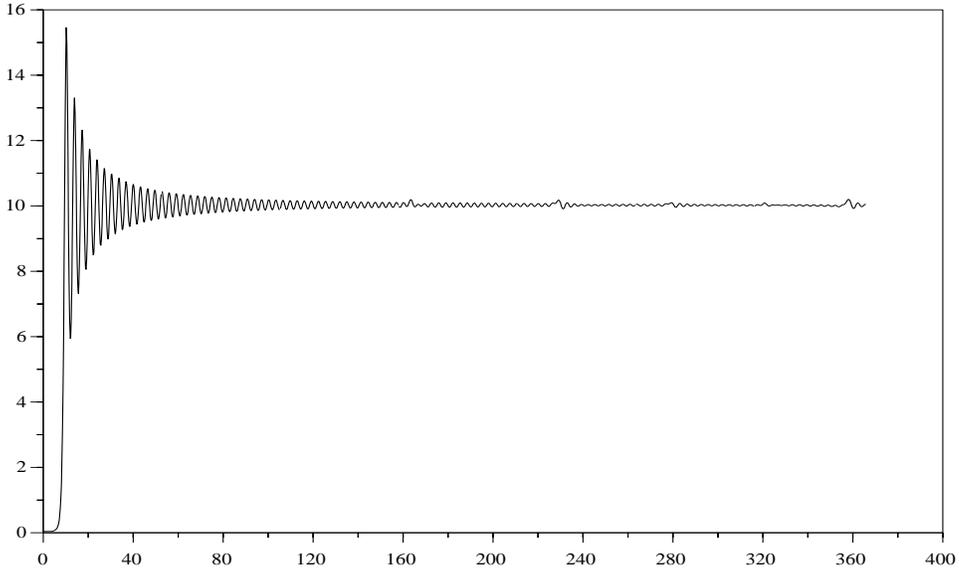,height=9cm,width=15cm}
\caption{\it The quantum back--reaction $\Sigma$, with the parameters as
in Fig. \ref{fig:mass2} }\label{fig:sigma}
\end{figure}

\begin{figure}
\epsfig{file=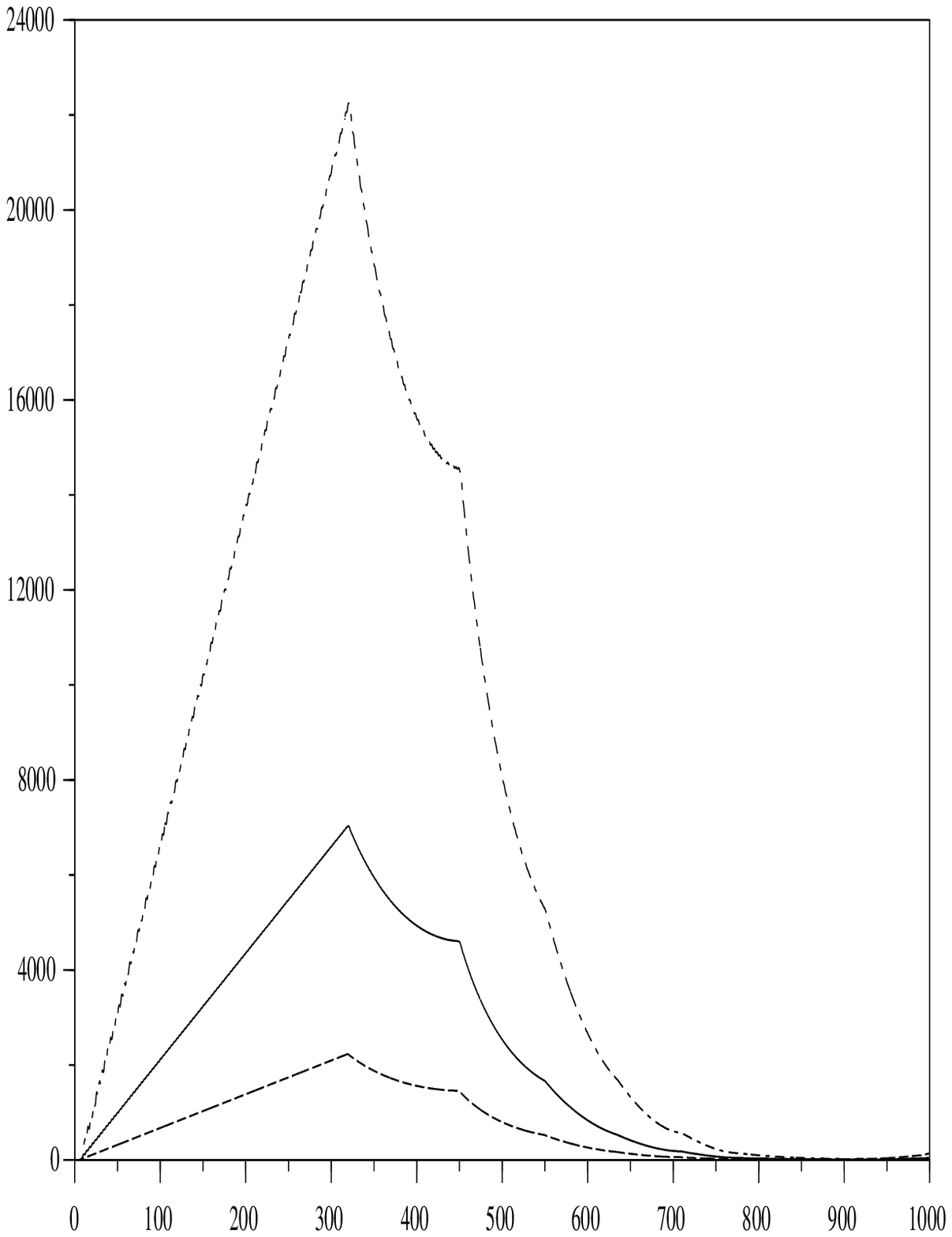,height=9cm,width=15cm}
\caption{\it Zero--mode amplitude evolution for different values of the
renormalized coupling constant $\l=0.01,0.1,1$, for $\frac{L}{2\pi}=100$ and
broken symmetry, with $\bar\phi=0$. }\label{fig:m0_l}
\end{figure}

\begin{figure}
\epsfig{file=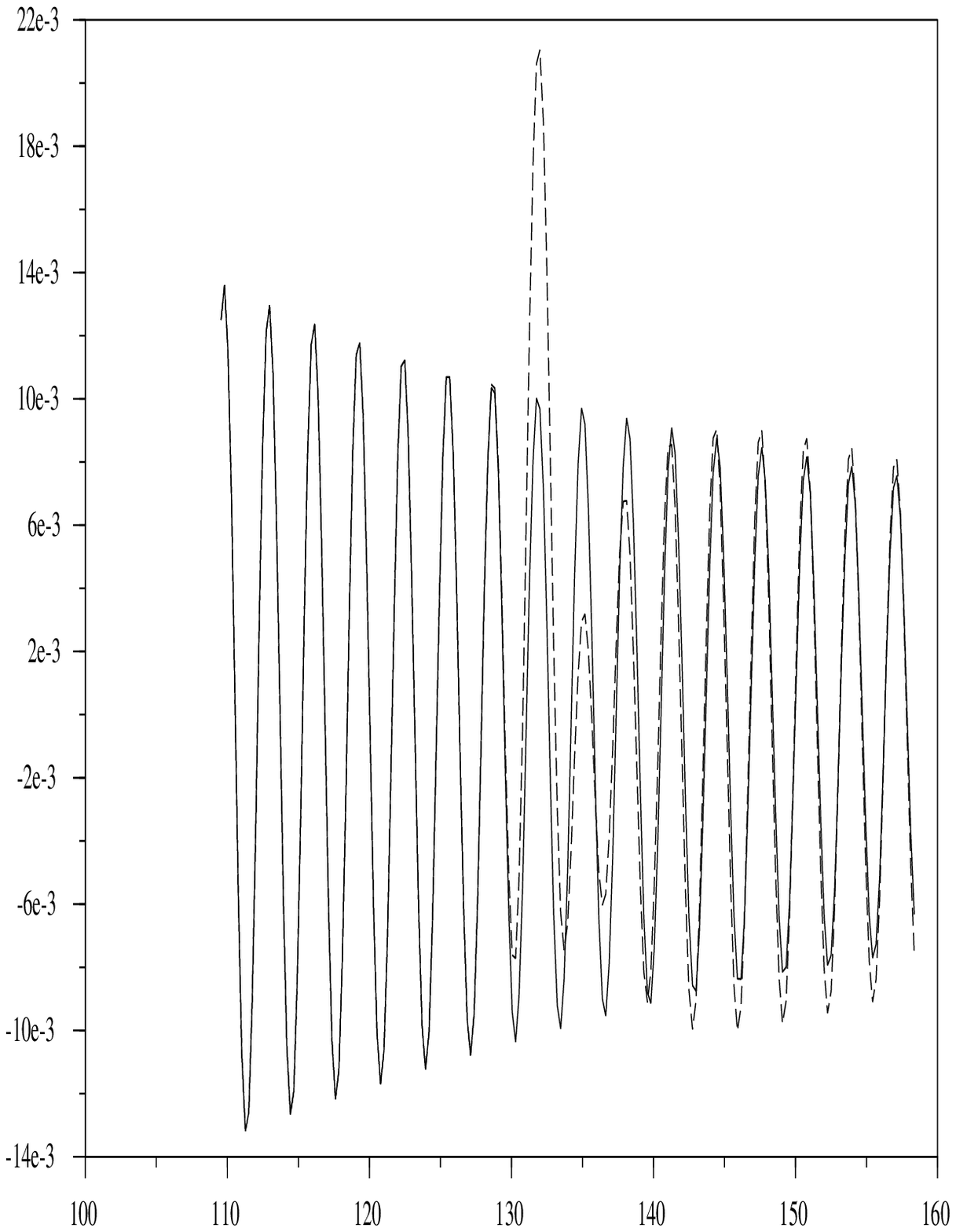,height=9cm,width=15cm}
\caption{\it Detail of $M^2$ near $t=\tau_L$ for $\frac{L}{2\pi}=40$ (dotted
line). The case $\frac{L}{2\pi}=80$ is plotted for comparison (solid line).}\label{fig:usc_mass}
\end{figure}

\begin{figure}
\epsfig{file=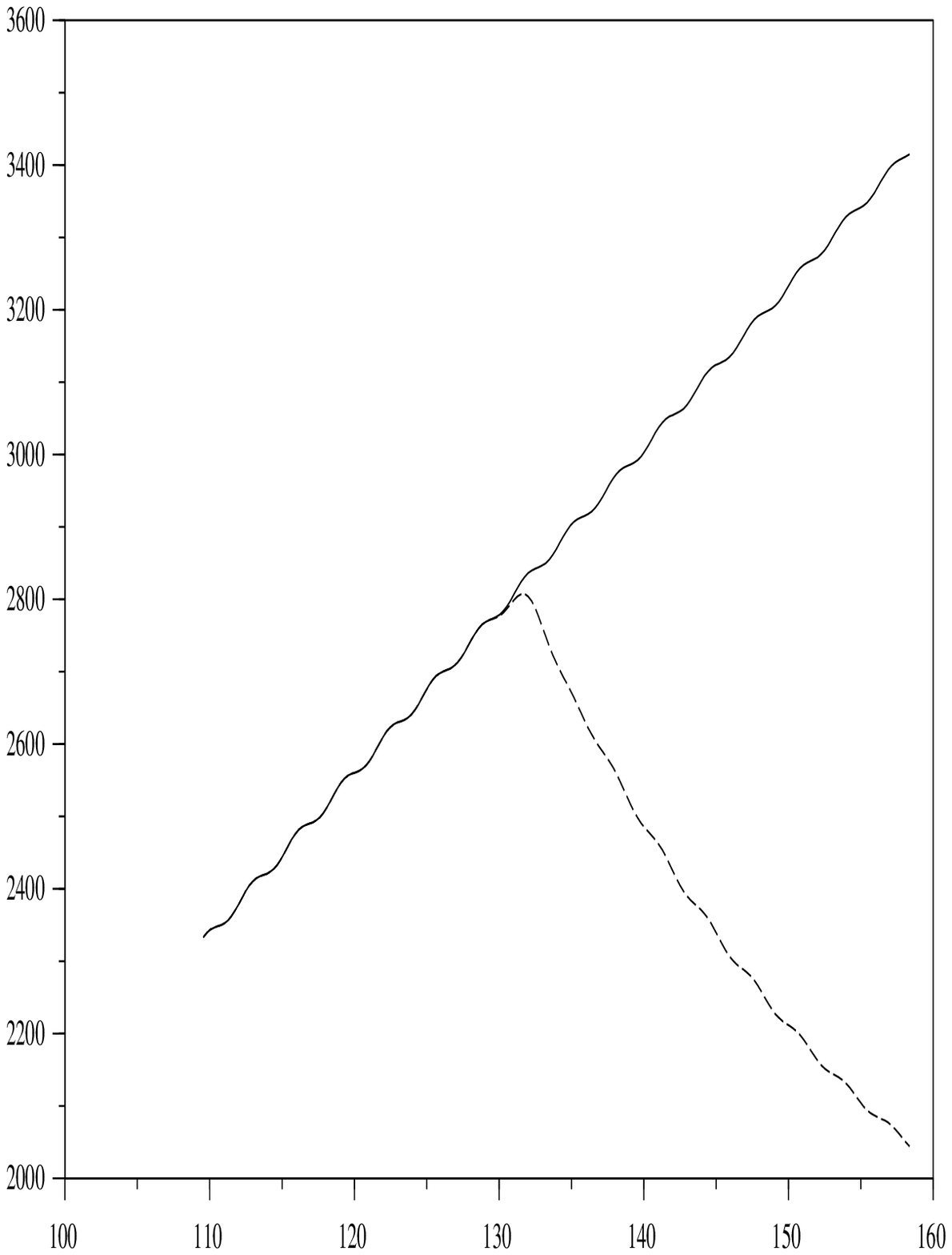,height=9cm,width=15cm}
\caption{\it Detail of $\s_0$ near $t=\tau_L$ for $\frac{L}{2\pi}=40$ (dotted
line). The case $\frac{L}{2\pi}=80$ is plotted for comparison (solid line).}\label{fig:usc_zm}
\end{figure}

The figures show the behavior of the width of the zero mode $\s_0$
(see Fig. \ref{fig:m0}), of the squared effective mass $M^2$ (see Fig.
\ref{fig:mass2} ) and of the back--reaction $\Sigma$ (see
Fig. \ref{fig:sigma}), in the more interesting case of broken
symmetry. The initial conditions are chosen in several different ways
(see the appendix for details), but correspond to a negative $M^2$ at
early times with the initial widths all microscopic, that is at most
of order $L^{1/2}$. This is particularly relevant for the zero--mode
width $\s_0$, which is instead macroscopic in the lowest energy state
when $\bar{\bds\phi}^2<v^2$, as discussed above. As for the
background, the figures are relative to the simplest case
$\bar{\bds\phi}= 0 =\dot{\bar{\bds\phi}}$, but we have considered also
initial conditions with $\bar{\bds\phi}> 0$, reproducing the
``dynamical Maxwell construction'' observed in ref. \cite{bdvhs}.  At
any rate, for the purposes of this work, above all it is important to
observe that, due to the quantum back--reaction, $M^2$ rapidly becomes
positive, within the so--called {\em spinodal time}
\cite{devega2,relax,bdvhs}, and then, for times before $\tau_L$, the
{\em weakly dissipative} regime takes place where $M^2$ oscillates
around zero with amplitude decreasing as $t^{-1}$ and a frequency
fixed by the largest spinodal wavevector, in complete agreement with
the infinite--volume results \cite{bdvhs}. Correspondingly, after the
exponential growth until the spinodal time, the width of the zero--mode
grows on average linearly with time, reaching a maximum for $t
\simeq\tau_L$. Precisely, $\s_0$ performs small amplitude oscillations
with the same frequency of $M^2$ around a linear function of the form
$A + B t$, where $A, B \approx \lambda ^{-1/2}$ (see
Fig. \ref{fig:m0_l}), confirming what already found in
refs. \cite{relax,bdvhs}; then quite suddenly it turns down and enters
long irregular Poincar\'e--like cycles. Since the spinodal oscillation
frequency does not depend appreciably on $L$, the curves of $\s_0$ at
different values of $L$ are practically identical for
$t<\tau_L$. After a certain number of complete oscillations, a number
that scales linearly with $L$, a small change in the behavior of $M^2$
(see Fig. \ref{fig:usc_mass}) determines an inversion in $\s_0$ (see
Fig. \ref{fig:usc_zm}), evidently because of a phase crossover between
the two oscillation patterns. Shortly after $\tau_L$ dissipation
practically stops as the oscillations of $M^2$ stop decreasing in
amplitude and become more and more irregular, reflecting the same
irregularity in the evolution of the widths.

We can give a straightforward physical interpretation for the presence
of the time scale $\tau_L$. As shown in \cite{bdvhs}, long after the
spinodal time $t_1$, the effective mass oscillates around zero with a
decreasing amplitude and affects the quantum fluctuations in such a
way that the equal--time two--point correlation function contains a
time--dependent non--perturbative disturbance growing at twice the
speed of light. This is interpreted in terms of large numbers of
Goldstone bosons equally produced at any point in space (due to
translation invariance) and radially propagating at the speed of
light.  This picture applies also at finite volumes, in the bulk, for
volumes large enough. Hence, due to our periodic boundary conditions,
after a time exactly equal to $L/2$ the forward wave front meets the
backward wave front at the opposite point with respect to the source,
and the propagating wave starts interfering with itself and heavily
changes the dynamics with respect to that in infinite volume. This
argument leads us to give the value of $\pi$ for the proportionality
coefficient between $\tau_L$ and $L/2\pi$, prevision very well
verified by the numerical results, as can be inferred by a look at the
figures.

The main consequence of this scenario is that the linear
growth of the zero--mode width at infinite volume should not be
interpreted as a standard form of Bose--Einstein Condensation (BEC),
occurring with time, but should be consistently considered as ``novel''
form of dynamical BEC, as found by the authors of \cite{bdvhs}. In
fact, if a macroscopic condensation were really there, the zero mode
would develop a $\delta$ function in infinite volume, that would be
announced by a width of the zero mode growing to values $O(L^{3/2})$
at any given size $L$. Now, while it is surely true that when we push
$L$ to infinity, also the time $\tau_L$ tends to infinity, allowing
the zero mode to grow indefinitely, it is also true that, at any fixed
though arbitrarily large volume, the zero mode never reaches a width
$O(L^{3/2})$, just because $\tau_L \propto L$. In other words, if we
start from initial conditions where $\s_0$ is microscopic, then it
never becomes macroscopic later on.

\begin{figure}
\epsfig{file=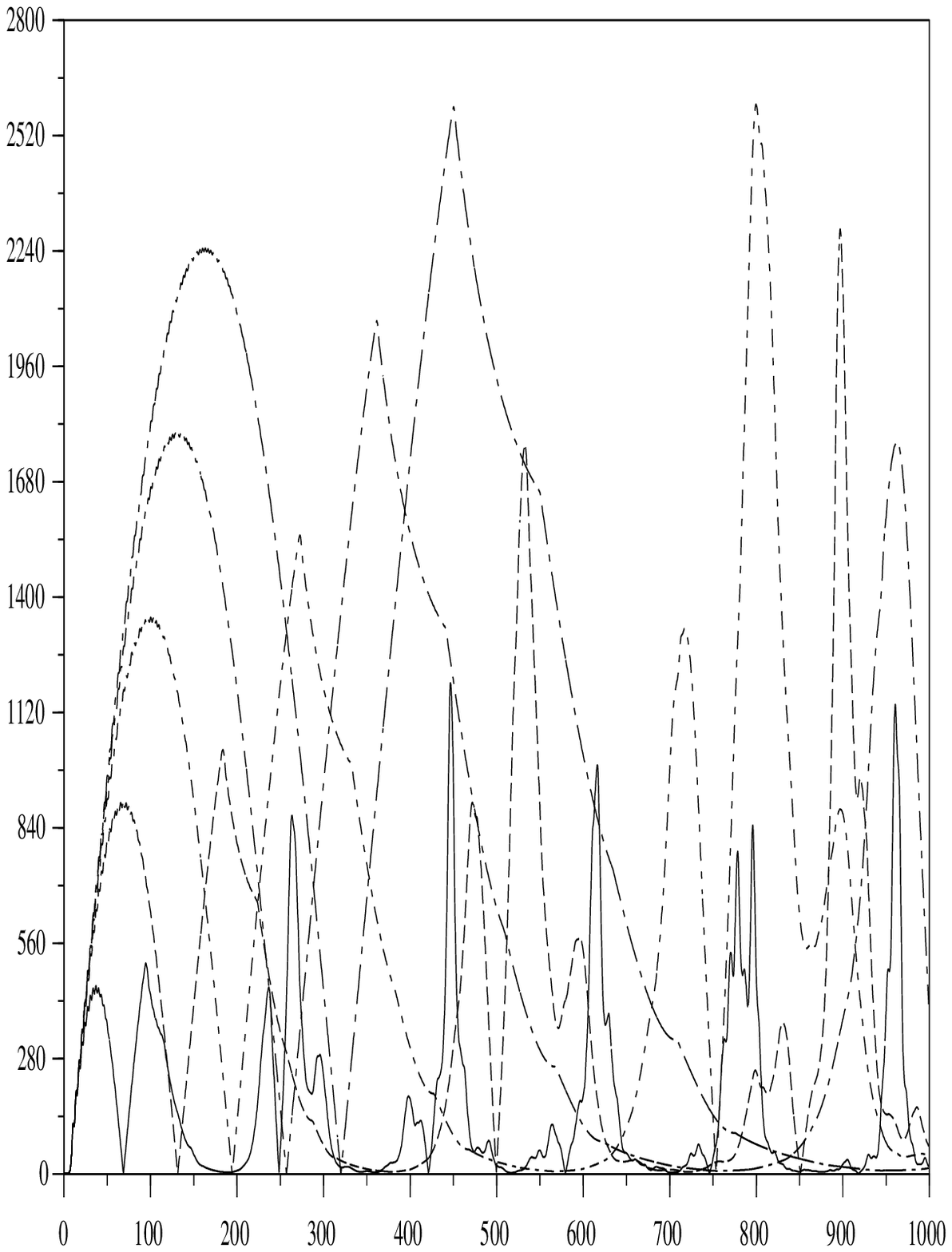,height=9cm,width=15cm}
\caption{\it Next--to--zero mode ($k=2\pi/L$) amplitude evolution for
different values of the size $\frac{L}{2\pi}=20,40,60,80,100$, for $\lambda =
0.1$ and broken symmetry, with $\bar\phi=0$.}\label{fig:m1}
\end{figure}

\begin{figure}
\epsfig{file=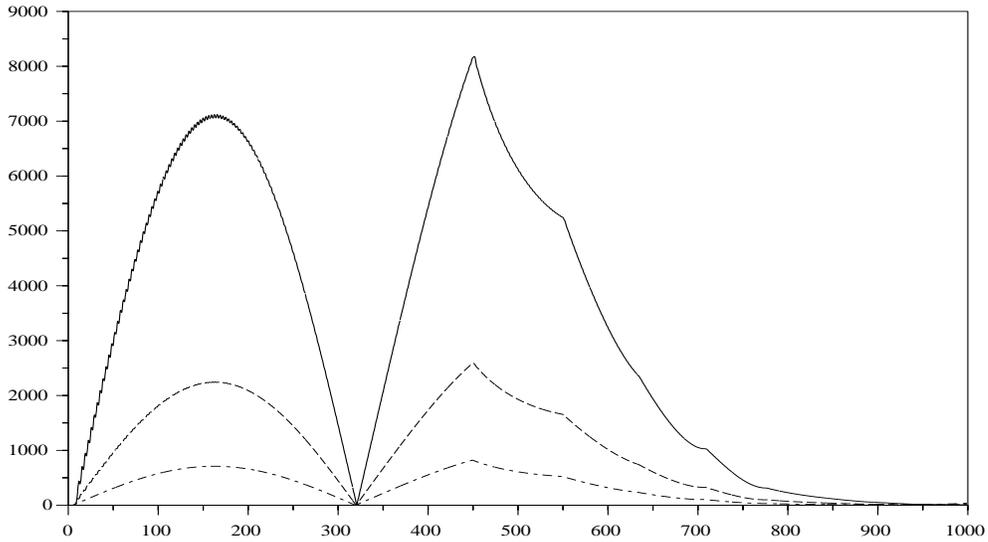,height=8.5cm,width=15cm}
\caption{\it Next--to--zero mode ($k=2\pi/L$) amplitude evolution for
different values of the renormalized coupling constant $\l=0.01,0.1,1$,
for $\frac{L}{2\pi}=100$ and broken symmetry, with $\bar\phi=0$.}\label{fig:m1_l}
\end{figure}

On the other hand, looking at the behavior of the mode functions of
momenta $k=(2\pi/L)n$ for $n$ fixed but for different values of $L$,
one realizes that they obey a scaling similar to that observed for the
zero--mode: they oscillate in time with an amplitude and a period that
are $O(L)$ (see fig. \ref{fig:m1} and \ref{fig:m1_l}). Thus, each mode shows a
behavior that is exactly half a way between a macroscopic amplitude
[i.e. $O(L^{3/2})$] and a usual microscopic one [i.e. at most
$O(L^{1/2})$]. This means that the spectrum of the quantum
fluctuations at times of the order of the diverging volume can be
interpreted as a {\em massless} spectrum of {\em interacting}
Goldstone modes, because their power spectrum develops in the limit a
$1/k^2$ singularity, rather than the $1/k$ pole typical of free
massless modes. As a consequence the equal--time field correlation
function [see eq. (\ref{2pN})] will fall off as $|\bds x-\bds y|^{-1}$
for large separations smaller only than the diverging elapsed time.
This is in accord with what found in \cite{bdvhs}, where the same
conclusion where reached after a study of the correlation function for
the scalar field in infinite volume.

The fact that each mode never becomes macroscopic, if it started
microscopic, might be regarded as a manifestation of unitarity in the
large $N$ approximation: an initial gaussian state with only
microscopic widths satisfies clustering and clustering cannot be
spoiled by a unitary time evolution. As a consequence, in the
infinite--volume late--time dynamics, the zero--mode width $\s_0$ does
not play any special role and only the behavior of $\s_k$ as $k\to 0$
is relevant. As already stated above, it turns out from our numerics
as well as from refs. \cite{relax,bdvhs,tsunami} that this behavior is
of a novel type characteristic both of the out--of--equilibrium
dynamics and of the equilibrium finite--temperature theory
\cite{simio}, with $\s_k \propto 1/k$.

A final comment should be made about the periodic boundary conditions
used for these simulations. This choice guarantees the translation
invariance of the dynamics needed to consider a stable uniform
background. If we had chosen other boundary conditions (Dirichlet or
Neumann, for instance), the translation symmetry would have been broken
and an uniform background would have become non-uniform pretty
soon. Of course, we expect the bulk behavior to be independent of the
particular choice for the boundary conditions in the infinite volume
limit, even if a rigorous proof of this statement is still lacking.

\section{Discussion and perspectives}\label{conclusion}
In this work we have presented a rather detailed study of the
dynamical evolution out of equilibrium, in finite volume (a cubic box
of size $L$ in $3$D), for the $\phi^4$ QFT in the large $N$ limit.
For comparison, we have also analyzed some static characteristics of
the theory both in unbroken and broken symmetry phases.

We have reached the conclusion, based on strong numerical evidence,
that the linear growth of the zero--mode quantum fluctuations,
observed already in the large $N$ approach of
refs. \cite{relax,bdvhs,tsunami}, may be consistently interpreted as a
``novel'' form of dynamical Bose--Einstein condensation, different
from the traditional one in finite temperature field theory at
equilibrium. In fact, in finite volume, $\s_0$ never grows to
$O(L^{3/2})$ if it starts from a microscopic value, that is at most of
order $L^{1/2}$. On the other hand all long--wavelength fluctuations
rapidly become of order $L$, signalling a novel infrared behavior
quite different from free massless fields at equilibrium [recall that
the large $N$ approximation is of mean field type, with no direct
interaction among particle excitations]. This is in agreement with the
properties of the two--point function determined in \cite{bdvhs}.

The numerical evidence for the linear dependence of $\tau_L$ on $L$ is
very strong, and the qualitative argument given in the previous
section clearly explains the physics that determines it. Nonetheless a
solid analytic understanding of the detailed (quantitative) mechanism
that produces the inversion of $\dot{\s}_0$ around $\tau_L$ and its
subsequent irregular behavior, is, at least in our opinion, more
difficult to obtain. One could use intuitive and generic arguments
like the quantization of momentum in multiples of $2\pi/L$, but the
evolution equations do not have any simple scaling behavior towards a
universal form, when mass dimensions are expressed in multiples of
$2\pi/L$ and time in multiples of $L$. Moreover, the qualitative form
of the evolution depends heavily on our choice of initial
conditions. In fact, before finite volumes effects show up, the
trajectories of the quantum modes are rather complex but regular
enough, having a small-scale quasi-periodic almost mode-independent
motion within a large-scale quasi-periodic mode-dependent envelope,
with a very delicate resonant equilibrium [Cfr. Fig. 1 and
7]. Apparently [Cfr. Fig. 5 and 6], it is a sudden small beat that
causes the turn around of the zero-mode and of the other low-lying
modes (with many thousands of coupled modes, it is very difficult for
the delicate resonant equilibrium to fully come back ever again), but
we think that a deeper comprehension of the non--linear coupled
dynamics is needed in order to venture into a true analytic
explanation.

On the other hand it is not difficult to understand why $\tau_L$ does
not depend appreciably on the coupling constant: when finite-volume
effects first come in, that is when the wave propagating at the speed
of light first starts to interfere with itself, the quantum
back-reaction $\lambda\Sigma$ has settled on values of order 1,
because the time $\tau_L \simeq L/2$ is much greater than the spinodal
time $t_1$. The slope of the linear envelope of the zero mode does
depend on $\lambda$ because it is fixed by the early exponential
growth. Similarly, it is easy to realize that the numerical
integrations of refs. \cite{relax,bdvhs,tsunami} over continuum
momenta correspond roughly to an effective volume much larger than
ours, so that the calculated evolution remained far away from the
onset of finite-volume effects.

The main limitation of the large $N$ approximation, as far as the
evolution of the widths $\s_k$ is concerned, is in its intrinsic
gaussian nature. In fact, one might envisage a scenario in which,
while gaussian fluctuations stay microscopic, non--gaussian
fluctuations grow in time to a macroscopic size. Therefore, in order
to clarify this point and go beyond the gaussian approximation, we are
going to consider, in a forthcoming work \cite{finvHF}, a
time--dependent HF approximation capable in principle of describing
the dynamics of non--gaussian fluctuation of a single scalar field
with $\phi^4$ interaction.

Another open question concerns the connection between the minima of
the effective potential and the asymptotic values for the evolution of
the background, within the simplest gaussian approximation. As already
pointed out in \cite{bdvhs}, a dynamical Maxwell construction occurs
for the $O(N)$ model in infinite volume and at leading order in $1/N$
in case of broken symmetry, in the sense that any value of the
background within the spinodal region can be obtained as large time
limit of the evolution starting from suitable initial conditions. It
would be very enlightening if we could prove this ``experimental''
result by first principles arguments, based on CTP
formalism. Furthermore, preliminary numerical evidence \cite{finvHF}
suggests that something similar occurs also in the Hartree
approximation for a single field, but a more thorough and detailed
analysis is needed.

It would be interesting also to study the dynamical realization of the
Goldstone paradigm, namely the asymptotic vanishing of the effective
mass in the broken symmetry phases, in different models; this issue
needs further study in the $2D$ case \cite{chkm}, where it is known
that the Goldstone theorem is not valid.

\section{Acknowledgements}
C. D. thanks D. Boyanovsky, H. de Vega, R. Holman and M. Simionato
for very interesting discussions. C. D. and E. M. thank MURST and INFN for
financial support. Part of the results contained in this work has been
presented by E. M. at the 1999 CRM Summer School on ``Theoretical
Physics at the End of the XXth Century'', held in Banff (Alberta),
Canada, June 27 - July 10, 1999. E. M. thanks CRM (Universit\'e de
Montr\`eal) for partial financial support.

\appendix
\section{Details of the numerical analysis}\label{num}
We present here the precise form of the evolution equations for the
field background and the quantum mode widths, which control the
out--of--equilibrium dynamics of the $\phi ^4$ model in finite volume
at the leading order in the $1/N$ approach, as described in sections
\ref{ooedN}. We restrict here our attention to the tridimensional
case.

Let us begin by noticing that each eigenvalue of the Laplacian operator in a
$3D$ finite volume is of the form $k_n ^2 = \left( \frac{2 \pi}{L}
\right) ^2 n$, where $n$ is a non--negative integer obtained as the sum of
three squared integers, $n = n _x ^2 + n _y ^2 + n _z ^2$. Then we
associate a degeneracy factor $g_n$ to
each eigenvalue, representing the
number of different ordered triples $(n _x, n _y, n _z)$ yielding the
same $n$. One may verify  that $g_n$ takes on the {\em continuum} value of
$4 \pi k ^2$ in the infinite volume limit, where $k=\left( \frac{2
\pi}{L} \right) ^2 n$ is kept fixed when $L \to \infty$.

Now, the system of coupled ordinary differential equations is, in case
of the large $N$ approach,
\begin{equation}\label{snumeq}
\left[ \frac{d ^2}{dt ^2} + M ^2\right] \phi
= 0 \;, \quad 
\left[ \frac{d ^2}{dt ^2} + \left( \frac{2 \pi}{L}
\right) ^2 n + M ^2 \right] \s_n - \frac1{4\s_n^3} =0
\end{equation}
where the index $n$ ranges from $0$ to ${\cal N}^2$, ${\cal N}=\Lambda
L/2\pi$ and $M^2(t)$ is defined by the eq. (\ref{Nunbgap}) in case of
unbroken symmetry and by eq. (\ref{Nbgap}) in case of broken symmetry.
The back--reaction $\Sigma$ reads, in the notations of this
appendix
\begin{equation*}
\Sigma = \frac{1}{L^D} \sum_{n=0}^{{\cal N}^2} g_n \s_n ^2 
\end{equation*}
Technically it is simpler to treat an equivalent set of
equations, which are formally linear and do not contain the singular
Heisenberg term $\propto \s_n^{-3}$. This is done by introducing the
complex mode amplitudes $z_n=\s_n\exp(i\t_n)$, where the phases $\t_n$
satisfy $\s_n^2\dot\t_n=1$. Then we find a discrete version of the
equations studied for instance in ref \cite{devega2},
\begin{equation}\label{numeq}
\left[ \frac{d ^2}{dt ^2} + \left( \frac{2 \pi}{L}
\right) ^2 n + M ^2 \right] z_n=0 \;,\quad 
\Sigma = \frac{1}{L^D} \sum_{n=0}^{{\cal N}^2} g_n |z_n| ^2
\end{equation}
subject to the Wronskian condition
\begin{equation*}
	z_n\,\dot{\bar{z_n}} - \bar{z_n}\,\dot z_n = -i
\end{equation*}
One realizes that the Heisenberg term in $\s_n$ corresponds to the 
centrifugal potential for the motion in the complex plane of $z_n$.

Let us now come back to the equations (\ref{numeq}). To solve these
evolution equations, we have to choose suitable initial conditions
respecting the Wronskian condition.  In case of unbroken symmetry,
once we have fixed the value of $\phi$ and its first time
derivative at initial time, the most natural way of fixing the initial
conditions for the $z_n$ is to require that they minimize the energy
at $t=0$. We can obviously fix the arbitrary phase in such a way to
have a real initial value for the complex mode functions
\begin{equation*}
z _n ( 0 ) = \frac{1}{\sqrt{2 \Omega _n}} \hspace{1 cm} \frac{d z _n}{dt} ( 0
) = \imath \sqrt{\frac{\Omega _n}{2}}
\end{equation*}
where $\Omega _n = \sqrt{k ^2 _n + M ^2 ( 0 )}$. The initial squared
effective mass $M ^2 (t= 0)$, has to be determined self-consistently,
by means of its definition (\ref{Nunbgap}).

In case of broken symmetry, the gap equation is a viable mean for
fixing the initial conditions only when $\phi$ lies outside the
spinodal region [see eq (\ref{NiceN})]; otherwise, the gap equation
does not admit a positive solution for the squared effective mass. In
that case, we have to resort to other methods, in order to choose the
initial conditions. Following the discussion presented in \ref{ooedN},
one possible choice is to set $\s_k^2 = \frac1{2\sqrt{k^2+|M^2|}}$ for
$k^2<|M^2|$ in eq. (\ref{newgap}) and then solve the corresponding gap
equation (\ref{newgap}). An other acceptable choice would be to solve
the gap equation (\ref{newgap}), once we have set a massless spectrum
for all the spinodal modes but the zero mode, which is started from an
arbitrary, albeit microscopic, value.

There is actually a third possibility, that is in some sense half a
way between the unbroken and broken symmetry case. We could allow for
a time dependent bare mass, in such a way to simulate a sort of {\em
cooling down} of the system. In order to do that, we could start with
a unbroken symmetry bare potential (which fixes initial conditions
naturally via the gap equation) and then turn to a broken symmetry one
after a short interval of time. This evolution is achieved by a proper
interpolation in time of the two inequivalent parameterizations of the
bare mass, eqs. (\ref{Nm2ren}) and (\ref{minimum}).

We looked for the influence this different choices could produce in
the results and indeed they depend very little and only
quantitatively from the choice of initial condition we make.

As far as the numerical algorithm is concerned, we used a $4$th order
Runge-Kutta algorithm to solve the coupled differential equations
(\ref{numeq}), performing the computations in boxes of linear size
ranging from $L = 20\pi$ to $L = 400\pi$ and verifying the conservation of
the Wronskian to order $10^{-5}$. Typically, we have chosen values of
$\cal N$ corresponding to the UV cutoff $\Lambda$ equal to small
multiples of $m$ for unbroken symmetry and of $v\sqrt{\l}$ for broken
symmetry. In fact, the dynamics is very weakly sensitive to the
presence of the ultraviolet modes, once the proper subtractions are
performed. This is because only the modes inside the unstable
(forbidden or spinodal) band grow exponentially fast, reaching soon
non perturbative amplitudes (i.e. $ \approx \lambda ^{-1/2}$), while
the modes lying outside the unstable band remains perturbative,
contributing very little to the quantum back--reaction \cite{relax} and
weakly affecting the overall dynamics. The unique precaution to take
is that the initial conditions be such that the unstable band lay well
within the cutoff.

\vskip 0.5 truecm

\vskip 1 truecm

\vskip 1 truecm

\vskip 1 truecm

\end{document}